\documentclass[11pt]{article}

\usepackage[preprint]{acl}

\usepackage{times}
\usepackage{latexsym}

\usepackage[T1]{fontenc}

\usepackage[utf8]{inputenc}

\usepackage{microtype}

\usepackage{inconsolata}

\usepackage{graphicx}

\usepackage{amsmath}
\usepackage{amssymb}
\usepackage{pgfplots}
\usepackage{subcaption}
\usepackage{multirow}
\usepackage{comment}
\usepackage{xcolor}
\usepackage{booktabs}

\title{Revisiting Audio-language Pretraining for Learning General-purpose Audio Representation}

\author{
 \textbf{Wei-Cheng Tseng\textsuperscript{1}$^\dagger$},
 \textbf{Xuanru Zhou\textsuperscript{2}$^\dagger$$^\ast$},
 \textbf{Mingyue Huo\textsuperscript{3}$^\dagger$$^\ast$},
 \textbf{Yiwen Shao\textsuperscript{4}$^\ddagger$},
 \textbf{Hao Zhang\textsuperscript{4}},
 \textbf{Dong Yu\textsuperscript{4}}
 \thanks{Equal Contribution.}
 \thanks{Work done during internship at Tencent AI Lab Seattle.}
 \thanks{Corresponding author: Yiwen Shao. Email: yiwenyshao@global.tencent.com}
\\
 \textsuperscript{1}University of Texas at Austin 
 \textsuperscript{2}Zhejiang University
 \\
 \textsuperscript{3}University of Illinois Urbana-Champaign
 \textsuperscript{4}Tencent AI Lab Seattle
\\
 \texttt{raytseng@utexas.edu, yiwenyshao@global.tencent.com}
}

\begin{document}
\maketitle
\begin{abstract}
Audio-language pretraining (ALP) holds promise for learning general-purpose audio representation, yet remains underexplored. 
Crucially, there is no consensus on whether audio–language models can build effective general-purpose audio encoders, nor a systematic understanding of how pretraining objectives behave across diverse tasks and scales.
We identify three key barriers: limited scale of audio-text corpora, limited coverage of audio attributes in existing caption corpora, and lack of systematic exploration and evaluation.
To fill this gap, we present the first principled empirical study of ALP.
We first introduce CaptionStew, a 10.7M caption dataset aggregating open-source audio-text corpora across multiple domains and captioning focuses.
We then conduct the first comprehensive evaluation comparing contrastive and captioning objectives for learning audio representation across speech, music, and environmental sound tasks.
Our results not only demonstrate that ALP yields competitive, transferable representations, but reveal critical trade-offs: contrastive learning offers superior data efficiency, while captioning exhibits better scalability.
Furthermore, we find that the benefits of supervised initialization often diminish at larger scales, challenging common practices.
By grounding these claims in empirical evidence, we establish a viable pathway toward general-purpose audio representation learning, guiding future research.
\end{abstract}

\pgfplotsset{compat=1.18}

\section{Introduction}
Representation learning has long been central to audio processing\footnote{In this work, audio processing refers to audio understanding, speech analysis and music understanding, while excluding automatic speech recognition}.
Current approaches are predominated by supervised learning~\citep{Panns, HTSAT, ecapa} and self-supervised learning~\citep{BEATs, wav2vec2, hsu2021hubert, MERT}, which have consistently enhanced performance across various speech and audio benchmarks~\citep{superb, hear, marble}.
Despite these advances, most existing methods are still optimized for relatively narrow task scopes rather than general-purpose use. models excelling at environmental sound classification, for example, often fail to capture speaker identity or paralinguistic attributes, and vice versa~\citep{hear}. 
Thus, learning audio representations that transfer robustly across diverse audio processing tasks remains an actively pursued and unresolved challenge.


A promising alternative is audio–language pretraining (ALP)~\citep{clap, LAION-CLAP}, which grounds audio perception in natural language descriptions (captions).
In this framework, text serves as a flexible semantic scaffold, enabling supervision spanning multiple levels of granularity, from coarse event categories (e.g., “dog barking,” “applause”) to fine-grained acoustic attributes (e.g., speaking style or musical structure), offering a unified path toward general audio understanding~\citep{MMAU,huangdynamic,airbench,ALPsurvey}.

The success of vision–language pretraining underscores this promise.
Models like CLIP~\citep{CLIP} and AIM-v2~\citep{AIMv2} not only power vision–language alignments, but also yield representations that transfer effectively to a broad range of downstream vision tasks~\citep{liu2023visual, minderer2022simple, crowson2022vqgan}.
For audio, however, analogous evidence remains limited.
Existing audio–language models~\cite{clap, LAION-CLAP, wavcaps, audiosetcaps} remain largely confined to retrieval tasks, leaving the community without a systematic understanding of whether ALP can serve as a practical route to general-purpose audio representation learning.
Fundamental questions remain unanswered: how do different pretraining objectives behave and scale, and how does transfer performance vary across heterogeneous audio tasks such as speaker identification and audio event classification?
The absence of empirical evidence regarding these questions has hindered progress and led to uncertainty of design choices.

We identify three key challenges that have constrained progress.
\textbf{First}, unlike vision–language learning, which benefits from web-scale image–text corpora containing billions of pairs~\citep{LAION5B, datacomp}, audio lacks comparably large open audio–text resources. Existing audio caption datasets typically remain at or below the million-pair scale~\citep{audiosetcaps, wavcaps, audiocaps, clotho}, fundamentally limiting the scaling potential of ALMs.
\textbf{Second}, current corpora offer limited semantic coverage: many captions describe only what sound events are present, while providing much less coverage of other important audio attributes such as speaker traits, musical properties, or acoustic environment.
This imbalanced focus limits the model’s ability to learn representations that capture the full range of audio semantics.
\textbf{Third}, prior ALP works have focused predominantly on contrastive learning and audio–text retrieval benchmarks.
Systematic studies on alternative pretraining objectives and comprehensive evaluations across a wide suite of audio understanding tasks remain scarce, limiting our understanding of what drives effective ALP.

In this work, we revisit ALP with the goal of reassessing its viability for learning general-purpose audio representation.
Rather than proposing a new model architecture, we provide \textbf{a foundational empirical study that fills the critical knowledge gap} described above, establishing a rigorous baseline to guide future research in accordance with scientific best practices.
We begin by aggregating diverse open-source audio caption datasets into a unified resource, \textbf{CaptionStew}, enabling analysis at substantially larger scales and with greater caption diversity than prior work.
Using this testbed, we conduct the first comprehensive evaluation of ALP across diverse downstream tasks and evaluation protocols, showing that it yields competitive and transferable representations across speech, music, and environmental audio domains. 
Through a controlled comparison between contrastive and captioning objectives, we reveal a consistent trade-off: contrastive learning exhibits superior data efficiency, while captioning demonstrates better scalability.
We further analyze key training factors—data scaling and supervised initialization—showing that not all tasks benefit uniformly from increased data, and that the gains from supervised initialization diminish at larger scales and for tasks beyond audio event classification, challenging common practices in the field.
Finally, we discuss how limited lexical diversity in existing caption datasets might constrain performance scaling on certain attributes, suggesting potential directions for improvement.

Taken together, our study reveals actionable insights that were previously undocumented for audio community and occasionally contradict trends from other modalities.
They establish ALP as a practical and competitive approach for learning general-purpose audio representations and highlight key factors for future progress. 
To facilitate further research, we release data, training and evaluation code, and pretrained models\footnote{https://github.com/AudenAI/Auden}.


\begin{figure*}[t]
    \centering
    \includegraphics[width=0.88\linewidth]{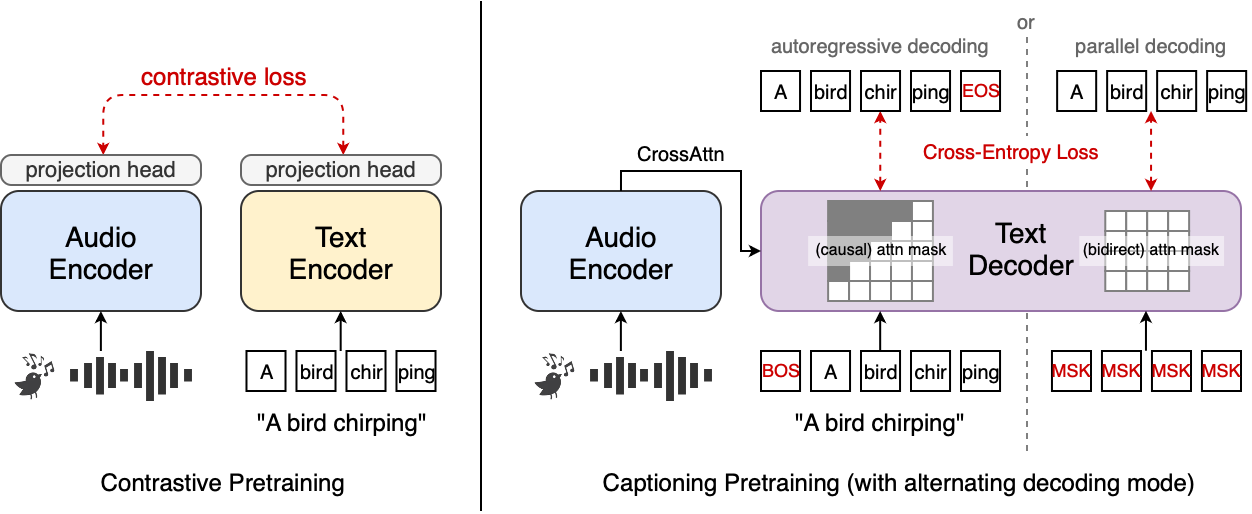}
    \caption{Audio-language pretraining objective studied in this work: contrastive and captioning.}
    \label{fig:placeholder}
\end{figure*}

\section{Related Work}
\textbf{Audio Representation Learning.}
Supervised models trained on labeled datasets have been fundamental to the field, including audio event classifiers~\citep{Panns, AST, HTSAT,  Dasheng}, speech recognition systems~\citep{whisper} and speaker recognition models~\citep{xvector, ecapa}.
These approaches remain widely adopted due to their strong performance on specific target tasks.
Self-supervised learning methods have also emerged, demonstrating benefits across speech~\citep{wav2vec2, hsu2021hubert, wavlm}, audio~\citep{audioMAE, BEATs, li2022atst}, and music~\citep{MERT,muq} without requiring labeled data.

\noindent\textbf{Audio–Language Pretraining.}
ALP has emerged as a promising approach for learning cross-modal representations. 
Most existing work focuses on contrastive objectives~\citep{clap, LAION-CLAP, wav2clip}, with recent extensions exploring combinations with other objectives~\citep{xu2023blat, zhu2024cacophony, niizumi2025m2d2}.
The field has also witnessed evolution in datasets, transitioning from human-annotated ones~\citep{audiocaps, clotho, MusicCaps} to recent LLM-augmented ones~\citep{wavcaps, audiosetcaps, fusionaudio, autoacd}, alongside domain-specific resources covering speaker characteristics~\citep{paraspeechcaps} and fine-grained musical attributes~\citep{jamendomaxcaps}.

\noindent\textbf{Universal Audio Understanding.}
The evaluation of audio understanding has evolved from task-specific benchmarks~\citep{superb, hear, marble} toward more complex evaluation framework.
Recent developments have emphasized LLM-based audio understanding systems~\citep{GAMA, LTU, midashenglm, AF3, chu2024qwen2, salmonn} that can handle natural langauge queries and complex reasoning tasks.
This shift has driven the development of corresponding evaluation benchmarks that assess models' abilities across diverse audio understanding scenarios~\citep{MMAU,airbench,huangdynamic,mmar}.
Our work contributes to this trend by providing the first comprehensive evaluation of ALP across discriminative tasks, audio-language alignment, and open-form question answering, bridging the gap between representation learning and universal audio understanding.

\section{Audio-language Pretraining}

ALP learns audio representations by establishing correspondence between audio and captions. The core concept is to leverage text as structured semantic supervision, enabling models to capture diverse information across speech, music, and environmental sounds within a unified framework.
ALMs typically employ a two-tower architecture: an audio encoder $f_{\text{a}}$ that maps raw audio signals into contextual representations, and a text component $f_{\text{t}}$ whose design depends on the training objective.
As shown in Figure~\ref{fig:placeholder}, we explore two complementary paradigms that differ fundamentally in how they establish audio-text correspondence: contrastive and captioning objective.
These two formulations reflect discriminative and generative perspectives of ALP, respectively, and provide a controlled basis for comparing objective-level trade-offs in general-purpose audio representation learning.

\subsection{Contrastive Objective}
Contrastive objective is proven to be a robust representation learning method~\citep{SimCLR, CLIP, wav2vec2} and have been a dominant approach for ALP~\citep{clap, LAION-CLAP, wav2clip}.
This approach aligns audio and text representations in a shared embedding space by maximizing similarity between paired samples while minimizing similarity between mismatched pairs.
Given a batch of paired samples $\{(a_i, t_i)\}_{i=1}^N$, the audio encoder produces frame- (or patch-) level representations that are pooled and projected to audio embeddings $\mathbf{z}^a_i$, while the text encoder $f_t$ generates corresponding text embeddings $\mathbf{z}^t_i$.
The symmetric InfoNCE loss~\citep{InfoNCE} is applied to optimize both modalities:
\begin{equation}
\begin{aligned}
\mathcal{L}_{\text{con}}=
&\frac{-1}{2N} \sum_{i=1}^N \Bigg[ 
\log \frac{\exp(\text{sim}(\mathbf{z}^a_i, \mathbf{z}^t_i)/\tau)}{\sum_{j=1}^N \exp(\text{sim}(\mathbf{z}^a_i, \mathbf{z}^a_j)/\tau)} \\
&+ 
\log \frac{\exp(\text{sim}(\mathbf{z}^t_i, \mathbf{z}^a_i)/\tau)}{\sum_{j=1}^N \exp(\text{sim}(\mathbf{z}^t_i, \mathbf{z}^a_j)/\tau)} \Bigg],
\end{aligned}
\end{equation}
where $\text{sim}(\cdot,\cdot)$ denotes cosine similarity and $\tau$ is a learnable temperature parameter. 
This objective encourages paired audio-text samples to be close in embedding space, encouraging semantic organization where similar content is grouped together.

\subsection{Captioning Objective}
Captioning objective takes a generative approach to audio-language alignment, learning representations by generating textual descriptions from audio.
We consider captioning to be a promising yet underexplored alternative in ALP, especially given the growing interest in general audio understanding systems that require interfaces to natural language.
Theoretically, the cross-attention mechanism provides frame-level supervision on the audio representation, offering denser learning signals than the utterance-level alignment used in contrastive learning.
Also, since captioning models the joint distribution over all caption tokens, it is inherently more sensitive to fine-grained attributes, relations, and word order, enabling richer relational grounding~\cite{ARO, SugarCrepe, tschannen2023image}.
Moreover, caption-based supervision is increasingly relevant given recent efforts toward general audio understanding systems~\citep{midashenglm,AF3}

Given an audio signal $a_i$, the encoder $f_a$ produces contextual representations $\mathbf{Z}^a_i$, which are fed into a transformer decoder $g_t$ through cross-attention. 
Inspired by the general encoder–decoder training recipe of CapPa~\citep{tschannen2023image}, we adopt a mixed decoding strategy that alternates between autoregressive decoding and parallel token prediction to enhance audio encoder representation learning.
In the autoregressive decoding, the decoder generates caption tokens $(y_1, \ldots, y_T)$ sequentially, with each token conditioned on the audio representation and previously generated tokens. Training follows the teacher-forcing approach with a cross-entropy loss:
\begin{equation}
\mathcal{L}_{\text{cap}} = - \sum_{t=1}^T \log p_\theta(y_t \mid y_{<t}, \mathbf{Z}^a_i),
\end{equation}

In parallel prediction, we replace the decoder input tokens with [MASK] tokens and remove the causal attention mask, forcing simultaneous prediction of all tokens based solely on audio features:
\begin{equation}
\mathcal{L}_{\text{par}} = - \sum_{t=1}^T \log p_\theta(y_t \mid \mathbf{Z}^a_i),
\end{equation}
This mode eliminates reliance on prior autoregressive context and forces each token prediction to depend solely on the audio representation, thereby strengthening encoder supervision.
We adopt mixed training in which a random fraction of each minibatch uses standard autoregressive decoding while the remainder uses parallel prediction. 
In our pilot ablation study, this mixed objective produced better downstream transfer than purely autoregressive decoding, so we use it as the captioning setup throughout the paper.


\section{CaptionStew Dataset}

\begin{table*}[!th]
\begin{tabular}{cc}
\begin{minipage}[t]{0.44\textwidth}
  \centering
  \captionof{table}{Comparison of publicly available audio caption datasets. The number of audio-text pairs (\#pair) and number of unique words (\#vocab) are shown here.}
  \label{tab:caption_datasets}
  \resizebox{1.0\textwidth}{!}{
  \renewcommand{\arraystretch}{0.86}
  \begin{tabular}{lrr}
\toprule
Audio Caption Dataset            & \#pair & \#vocab \\
\midrule
\textit{  Human-annotated} & &\\
AudioCaps~\cite{audiocaps}          & 46K               & 4,844         \\
Clotho~\cite{clotho}             & 5K                & 4,366         \\
MusicCaps~\cite{MusicCaps}          & 5K                & 3,730         \\
\midrule
\textit{  LLM-augmented} & &\\
WavCaps~\cite{wavcaps}            & 403K              & 18,372        \\
AudioSetCaps~\cite{audiosetcaps}       & 1.9M              & 21,783        \\
FusionAudio~\cite{fusionaudio}        & 1.2M              & 18,403        \\
AutoACD~\cite{autoacd}            & 1.5M              & 20,491        \\
\midrule
CaptionStew (Ours) & 10.7M             & 56,586      \\
\bottomrule
\end{tabular}
}
\end{minipage}
&
\begin{minipage}[t]{0.50\textwidth}
  \centering
  \captionof{table}{Datasets used for evaluating  linear probing, audio-language task and open-form question answering performance (separated by lines). All metrics are higher the better. $^\dagger$reported with AIR-Bench~\cite{airbench}.}
  \label{tab:evaluation_tasks}
\resizebox{1.0\textwidth}{!}{
\renewcommand{\arraystretch}{0.86}
  \begin{tabular}{lll}
\toprule  
Evaluation Dataset & Task                                 & Metrics \\
\midrule
FSD-50k            & Multi-label audio event classification  & mAP     \\
VggSound           & Single-label audio event classification & accuracy    \\
VoxCeleb2          & Speaker identification                  & accuracy    \\
CREMA            & Speech emotion recognition              & accuracy   \\
MagnaTagATune      & Music tagging                           & mAP     \\
NSynth             & Musical instrument classification       & accuracy  \\
AS-strong    & Sound event detection & PSDS1 \\
\midrule
AudioCaps             & \multirow{3}{*}{\begin{tabular}[c]{@{}l@{}}Text-to-audio retrieval\\ Audio captioning\end{tabular}}      & \multirow{3}{*}{\begin{tabular}[c]{@{}l@{}}Recall@1\\ RougeL\end{tabular}}   \\
ParaSpeechCaps             &        &    \\
MusicCaps             &        &    \\
\midrule
ClothoAQA & \multirow{3}{*}{Open-formed question answering}& \multirow{3}{*}{Score$^\dagger$}\\
ParaLMQA & & \\
MusicQA & & \\
\bottomrule
\end{tabular}
}

\end{minipage}
\\
\end{tabular}
\end{table*}

To investigate the potential of audio–language pretraining for general-purpose representation learning, we collect a large-scale and diverse audio caption dataset that addresses key limitations in existing corpora: limited scale and limited semantic coverage. 
Audio signals inherently encode information across multiple dimensions—timbre, pitch, rhythm, semantic events, emotional tone, and acoustic environment—each amenable to different linguistic descriptions. 
However, existing large-scale audio caption datasets typically rely on a single caption-generation pipeline (Appendix~\ref{app:dataset_detail}), where all captions are produced through the same procedure—either human annotation following uniform guidelines or LLM-based synthesis—and therefore tend to exhibit a relatively narrow descriptive focus.
This uniformity offers consistency and scalability but introduces systematic stylistic biases and restricts lexical diversity.
Moreover, single-pipeline captions tend to exhibit limited semantic coverage and a narrow descriptive focus on only a subset of audio characteristics, often overlooking complementary acoustic attributes.

To fully leverage text as a flexible semantic scaffold for diverse audio representation learning, we embrace caption diversity across sources, styles, and descriptive granularities. 
Rather than creating captions through a single pipeline, we aggregate existing open-source corpora~\citep{audiocaps, clotho, MusicCaps, wavcaps, fusionaudio, audiosetcaps, paraspeechcaps, jamendomaxcaps}. 
These datasets span multiple audio domains—general sound events, expressive speech, and musical performance—and employ fundamentally different caption creation methodologies.
This aggregation yields captions that describe complementary aspects of audio with different levels of granularity, ranging from coarse event categories to finer-grained acoustic and stylistic attributes.
Please refer to Appendix~\ref{app:dataset_detail} for detail and examples of each source dataset.
The resulting dataset, \textbf{CaptionStew} (denoted by CS10M), contains 9.3 million audio samples paired with 10.7 million captions, spanning 37,290 hours across speech, music, and environmental domains. Compared with prior public audio caption resources, CaptionStew substantially increases scale while also broadening semantic coverage.
This makes it a useful and reproducible testbed for studying how audio–language pretraining behaves across objectives, tasks, and training scales.
Table~\ref{tab:caption_datasets} presents a comparison with existing audio caption datasets.

\section{Experimental Setup}
\subsection{Implementation Details}
\label{sec:implementation}

We pretrain all models on CaptionStew. The audio encoder uses a Zipformer-M architecture~\citep{Zipformer}, chosen for its efficiency on long sequences and fast convergence.
For contrastive pretraining, the text encoder follows BERT-base architecture~\citep{devlin2019bert}.
For captioning pretraining, the text decoder adopts the BART-base decoder architecture ~\citep{lewis2019bart}. 
We use twice as many encoder layers (12) as decoder layers (6) to ensure comparable training speed across objectives. 
We experiment with two scenarios: training from scratch (\textit{-scratch}) or initialized from pretrained checkpoints (\textit{-init}), following prior works in ALP~\citep{LAION-CLAP, wavcaps,audiosetcaps}.
Please refer to Appendix~\ref{app:full_implementation_details} for the full implementation details.

\subsection{Evaluation Protocols and Datasets}
\label{sec:eval}
We evaluate pretrained audio encoders across three protocols assessing discriminative capabilities, audio-language alignment, and open-formed question answering.
All experiments probe frozen representations from the audio encoder's final layer to ensure fair comparison. Table~\ref{tab:evaluation_tasks} and Appendix~\ref{app:eval_datasets} details the datasets and task metrics.

\noindent\textbf{Linear Probing} trains simple linear classifier on frozen representations.
We evaluate across a diverse set of tasks across audio domains, including audio event classification (AEC)~\citep{fsd50k, vggsound}, sound event detection (SED)~\citep{asstrong}, speaker identification (SID)~\citep{voxceleb2}, speech emotion recognition (SER)~\citep{cao2014crema}, music tagging (MTAG)~\citep{MTT} and musical instrument classification (INST)~\citep{NSynth}.

\noindent\textbf{Audio-language Alignments}
follow the LiT protocol~\citep{LiT}, adapting either pretrained text encoder~\citep{liu2019roberta} or text decoder~\citep{lewis2019bart} to align with frozen audio representations for performing retrieval and captioning tasks. 
We evaluate on audio-caption datasets spanning diverse domains: AudioCaps (AC)~\citep{audiocaps} for general sound event descriptions; ParaSpeechCaps (PSC)~\citep{paraspeechcaps} for speaking-style and acoustic-environment descriptions; and MusicCaps (MC)~\citep{MusicCaps} for fine-grained musical descriptions.

\noindent\textbf{Open-formed Question Answering}.
Acknowledging the trend of combining audio encoders with large language models (LLMs) for general audio understanding~\citep{GAMA, LTU}, we connects frozen audio encoders to a LLM (Qwen2.5-7B-Instruct~\citet{qwen25}) through lightweight adaptors.
We train only the adaptor on multiple audio QA datasets that span distinct domains: sound event understanding~\citep{clothoaqa}, speaker-related and paralinguistic understanding~\citep{huo2025auden}, and music understanding~\citep{mullama}. Evaluation is conducted on the corresponding tracks (sound, speaker-related, music; see Appendix~\ref{app:eval_datasets}) of AIR-Bench~\citep{airbench}.

\newcommand{\twolineheader}[2]{\shortstack{\textbf{#1}\\ \scriptsize #2}}

\begin{table*}[!h]
\caption{Evaluation results across tasks and protocols. $^\dagger$numbers quoted from other papers with consistent evaluation setup. $^\ddagger$state-of-the-art results on each task without any training constraints (e.g. full-finetuning) (see Appendix~\ref{app:main}). $^{\dagger\dagger}$no available prior work. $^{\ddagger\ddagger}$results of speaker emotion recognition, gender recognition, and age prediction in AIR-Bench~\cite{airbench}, respectively.}
\label{tab:main}
\begin{subtable}{\textwidth}
  \centering
  \caption{Linear Probing (with mean pooling)} 
\resizebox{\textwidth}{!}{
\renewcommand{\arraystretch}{0.84}
\begin{tabular}{lccccccccc}
\toprule
\multirow{3}{*}{Method} & \multirow{3}{*}{\shortstack{Model\\Initialization}}& \multirow{3}{*}{\shortstack{Audio-lang.\\Pretraining}}& \multicolumn{7}{c}{\textbf{linear probing}}                                    \\
\cmidrule{4-10}
                  & & & \twolineheader{AEC}{FSD50k} & \twolineheader{AEC}{VggSound} & \twolineheader{SID}{VoxCeleb2} & \twolineheader{SER}{CREMA} & \twolineheader{MTAG}{MagnaTagATune} & \twolineheader{INST}{NSynth} & \twolineheader{SED}{AS-Strong} \\
\midrule
\textbf{\textit{   Existing SSL Models}} & & & & \\
BEATs~\cite{BEATs}             &    SSL  &  -- & 0.565$^\dagger$   & --       & --        & -- & 0.400$^\dagger$      & \underline{75.90}$^\dagger$  & 0.034$^\dagger$     \\
Wav2vec 2.0~\cite{wav2vec2}       &    SSL   & -- & 0.342$^\dagger$   & --       & \underline{51.60}     & 56.10 & 0.317$^\dagger$      & 40.20$^\dagger$     & --        \\
MERT~\cite{MERT}              &    SSL   &  -- & --      & --       & --        & --    & 0.402$^\dagger$   & 72.60$^\dagger$  & --    \\
\midrule
\textbf{\textit{   Our Supervised Baselines}} & & & & \\
Zipformer-AEC~\cite{Zipformer}        & AudioSet SL  &  --  & 0.656   & \underline{56.46}    & 18.84     & 67.14 & 0.407   & 67.19  & \underline{0.216}     \\
\midrule
\textbf{\textit{   Our Audio-lang. Pretrained}} & & & & \\
Contrastive-\textit{scratch}       &     --     & CS10M & 0.625   & 50.87    & \textbf{46.67}     & 67.71 & 0.406   & 67.30  & 0.132     \\
Captioning-\textit{scratch}        &    --      & CS10M & 0.580   & 47.79    & 33.43     & 63.60 & 0.401   & 63.10  & 0.124     \\
Contrastive-\textit{init}       & AudioSet SL      & CS10M & \underline{\textbf{0.664}}   & \textbf{54.70}    & 38.17     & \underline{\textbf{68.84}} & 0.406   & \textbf{69.38}  & \textbf{0.187}     \\
Captioning-\textit{init}        & AudioSet SL      & CS10M & 0.652   & 53.13    & 26.23     & 65.86 & \underline{\textbf{0.410}}   & 67.16  & 0.145     \\
\midrule
\midrule
SOTA$^\ddagger$ & & & 0.655 & 59.50 & 96.20 & --$^{\dagger\dagger}$ & 0.414 & 79.20 & 0.374\\
\bottomrule
\end{tabular}
}
\end{subtable}

\vspace{1.0em} 
\begin{subtable}{\textwidth}
  \centering
  \caption{Audio-language Alignment / Open-form QA}
\resizebox{\textwidth}{!}{
\renewcommand{\arraystretch}{0.83}
\begin{tabular}{lccccccccc}
\toprule
        \multirow{2}{*}{Method}          & \multicolumn{3}{c}{\textbf{Captioning}} & \multicolumn{3}{c}{\textbf{Retrieval}} & \multicolumn{3}{c}{\textbf{Open-formed QA}}                          \\
\cmidrule(lr{0.5em}){2-4} \cmidrule(lr{0.5em}){5-7} \cmidrule(lr{0.5em}){8-10}
                  & AC    & PSC    & MC    & AC    & PSC    & MC        & Sound & Speaker-related$^{\ddagger\ddagger}$      & Music \\
\midrule
\textbf{\textit{   Our Supervised Baselines}} & & & & \\
Zipformer-AEC~\cite{Zipformer}        & 46.7          & 45.5              & \underline{\textbf{22.9}}           & 40.5         & 49.2              & 24.6           & 7.01           & 36.5 / 46.2 / 37.2 & 5.61             \\
\midrule
\textbf{\textit{   Our Audio-lang. Pretrained}} & & & & \\
Contrastive-\textit{scratch}       & 46.6          & 46.3              & 22.1           & 39.3         & \underline{\textbf{63.2}}              & 27.4           & 6.65           & 37.9 / \underline{\textbf{81.3}} / 63.4 & 5.86             \\
Captioning-\textit{scratch}           & 46.7          & \underline{\textbf{46.5}}              & \underline{\textbf{22.9}}           & 36.9         & 60.2              & 23.0           & 6.69           & \underline{\textbf{44.2}} / 65.4 / \underline{\textbf{69.0}} & \underline{\textbf{5.97}}             \\
Contrastive-\textit{init}          & \underline{\textbf{47.2}}          & 46.2              & 22.5           & \underline{\textbf{42.8}}         & 60.6              & \underline{\textbf{29.4}}           & 6.73           & 35.1 / 67.3 / 64.5 & 5.63             \\
Captioning-\textit{init}         & \underline{\textbf{47.2}}          & 45.9              & 22.6           & 42.2         & 55                & 28.2           & \underline{\textbf{7.06}}           & 32.4 / 49.5 / 45.6 & 5.50             \\
\midrule
\midrule
SOTA$^\ddagger$              & 52.2          & --$^{\dagger\dagger}$                & 26.2          & 44.4         & --$^{\dagger\dagger}$                & --$^{\dagger\dagger}$         & 6.99           & 60.0 / 82.5 / 62.4 & 6.79     \\
\bottomrule
\end{tabular}
}
\end{subtable}
\end{table*}

\subsection{Baseline Methods}
Recognizing the broad adoption of pretrained audio event classifiers in transfer learning~\citep{alonso2023efficient, cappellazzo2024parameter}, audio-language modeling~\citep{clap, LAION-CLAP} and general audio understanding~\citep{LTU, GAMA, midashenglm}, we select our pretrained Zipformer-based audio event classifier (denoted by Zipformer-AEC, described in Appendix~\ref{app:full_implementation_details}) as the primary baseline.
We also compare against representative self-supervised learning (SSL) models, each specialized for particular audio domains: BEATs~\citep{BEATs} for enviromental sound (or general audio); Wav2vec 2.0~\citep{wav2vec2} for speech signal; and MERT for music pieces.
Together, these baselines provide a broad comparative context for studying ALP toward general-purpose audio representation.


\section{Experiment Results}
We present our evaluation results in Table~\ref{tab:main}. Our analysis reveals key insights about objective design, representation quality, and the role of initialization.

\noindent\textbf{Contrastive vs. Captioning Objectives.} 
The two pretraining paradigms exhibit complementary strengths across evaluation protocols.
On linear probing tasks, contrastive learning consistently outperforms captioning, particularly excelling at audio event classification and speaker identification.
However, it is worth noting that this gap narrows substantially when the classifier learns to aggregate information across frames through multi-head attention pooling (Appendix~\ref{app:main}).
This observation reflects the objectives' inherent designs: contrastive learning explicitly optimizes for linearly separable clip-level representations, while captioning relies on cross-attention mechanisms over frame-level representations for text sequence generation.
This finding aligns with recent work highlighting how downstream module choices significantly impact the assessment of audio representation quality~\citep{speechbenchmarking}.
For language-involved tasks, both objectives demonstrate competitive performance, with captioning showing slight advantages in open-form question answering across multiple domains.
This suggests captioning's potential for language-involved audio understanding tasks.

\noindent\textbf{Impact of Supervised Initialization.} Initializing from supervised pretraining (AS SL) provides substantial benefits across most tasks, with notable improvements on audio event classification, sound event detection and audio-text retrieval.
The gains are particularly pronounced for contrastive objectives, suggesting that supervised pretraining provides useful inductive biases for contrastive learning.
However, these benefits diminish (or disappear entirely) when the attributes required for downstream tasks diverge from AudioSet's ontology. 
On speaker identification and music tagging, scratch-trained models often match or exceed initialized variants, indicating that AudioSet's focus on distinguishing between sound categories may bias representations toward event-level semantics rather than the speaker characteristics (voice timbre, speaking style) or musical structure (chord, rhythm) essential for these tasks.
These findings challenge common initialization practices for ALP and suggest the need for tailored pretraining strategies when targeting general-purpose audio representation learning.

\noindent\textbf{Competitive Performance Across Domains.} Our audio-language representations achieve strong transferability across diverse audio domains.
Compared to supervised baselines (Zipformer-AEC), our overall best-performing model (Contrastive-init) demonstrates superior performance on speaker identification, music understanding and audio-text retrieval while maintaining competitiveness on audio-event classification.
Against domain-specialized SSL methods (BEATs, Wav2vec 2.0, MERT), our approach consistently shows competitive performance.
This cross-domain performance validates our hypothesis that diverse caption aggregation enables broadly transferable representations, establishing ALP as a viable path toward learning general-purpose audio representation.

\begin{figure*}
    \centering
    \includegraphics[width=0.95\linewidth]{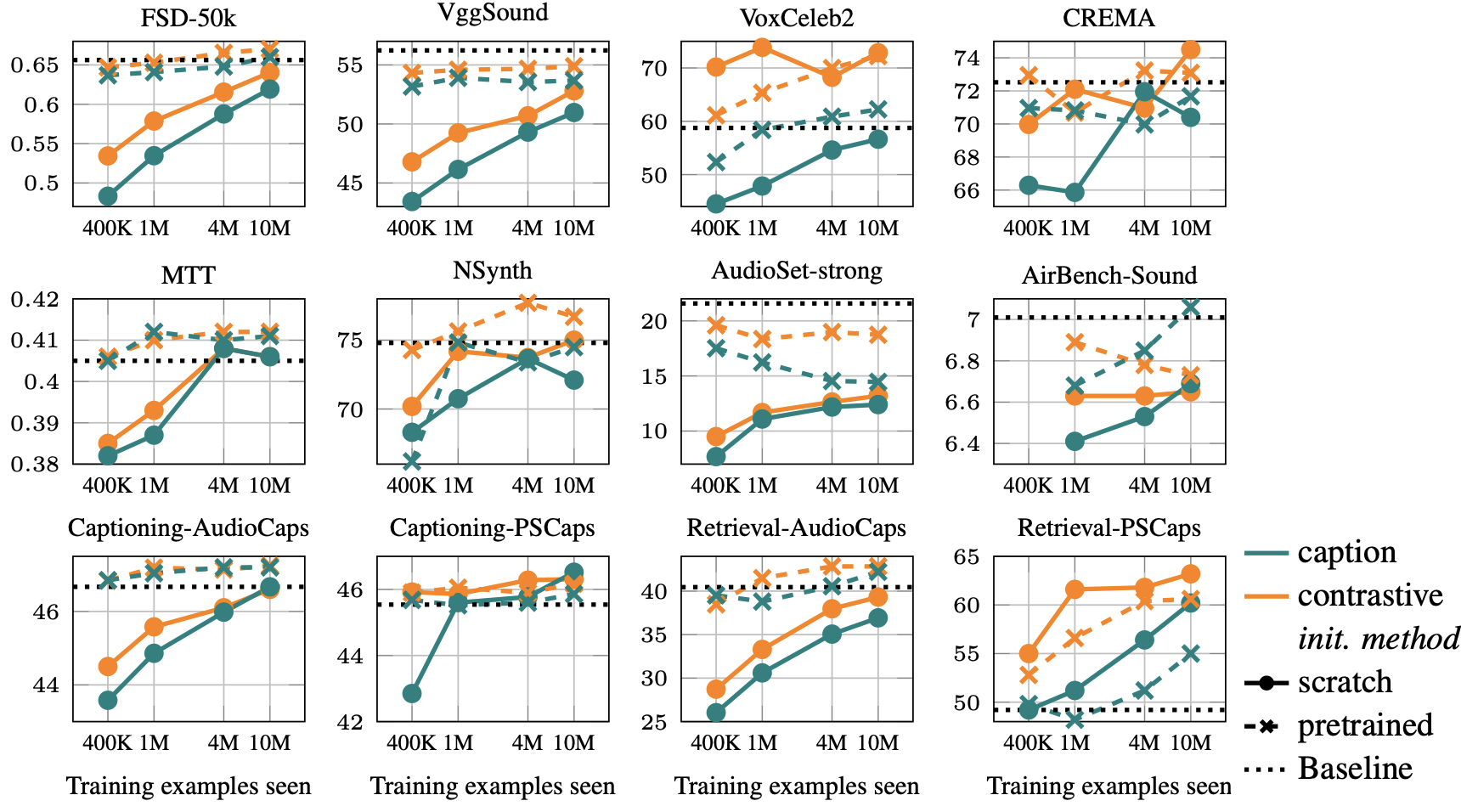}
    \caption{Data scaling behavior of contrastive vs. captioning objectives across representative tasks.}
    \label{fig:data_scaling}
\end{figure*}
\subsection{Data-Scaling Experiments}
To understand the scalability of audio–language pretraining, we conduct controlled experiments using CaptionStew subsets at 400K, 1M, 4M, and 10M (whole corpus) audio-text pairs. 
Importantly, these subsets are constructed in a strictly nested manner, i.e., $\mathcal{D}_{400k} \subset \mathcal{D}_{1M}\subset \mathcal{D}_{4M}\subset \mathcal{D}_{10M}$
, rather than via independent resampling.
Figure~\ref{fig:data_scaling} reveals distinct scaling patterns across objectives and evaluation protocols.

\noindent\textbf{Scaling Patterns.} Most tasks demonstrate consistent performance improvements with increased data scale, validating the potential of large-scale ALP. 
However, notable exceptions emerge that reveal fundamental limitations of current approaches.
Sound event detection, particularly for models initialized with AudioSet pretraining, exhibits a reverse scaling trend where performance degrades with more caption data. 
This suggests a potential conflict between natural language supervision--which typically describes audio characteristics and attributes--and temporal localization tasks requiring precise event boundaries.
Additionally, emotion recognition and instrument classification show weaker scaling gains compared to other tasks, likely reflecting limited caption diversity for these specific attributes in existing corpora (see Sec.~\ref{sec:lexical_analysis}).

\noindent\textbf{Contrastive vs. Captioning Scaling. }
Contrastive learning consistently outperforms captioning at varying data scales, particularly under less data and on discriminative tasks such as audio event classification. 
However, captioning demonstrates slightly better scaling properties, with distinct patterns emerging across task categories. 
or language-involved tasks--especially captioning and question answering--captioning matches or surpasses contrastive learning at our current 10M-pair scale.
On linear probing benchmarks, the gap remains substantial, with scaling trends suggesting captioning would require hundreds of millions of pairs to achieve parity with contrastive methods.

\noindent\textbf{Impact of Initialization at Scale.}
AudioSet initialization provides immediate performance gains but introduces diminishing returns at larger scales. Both contrastive learning and captioning show decreasing benefits from initialization as data scale increases, with scratch and initialized models achieving matched performance at larger scales on some tasks.
This suggests that pretrained initialization effectively bootstraps learning at small scales but may constrain the model's ability to adapt to the broader semantic space covered by large-scale caption data, potentially due to mismatch between AudioSet's ontology and diverse audio descriptions.

Overall, these findings reveal complementary behaviors: contrastive pretraining achieves superior data efficiency at current scales, while captioning shows better scalability, especially for language-involved tasks. 
Importantly, the diminishing returns of initialization at scale indicate that large-scale caption data can provide sufficient semantic supervision independent of domain-specific pretraining, challenging current practices of ALP and opening possibilities for learning general-purpose representations from diverse captions alone.

\subsection{Dataset Analysis}
\label{sec:lexical_analysis}

To understand the linguistic characteristics of CaptionStew, we analyze both its semantic coverage and its lexical diversity across constituent datasets.
Figure~\ref{fig:tsne} provides qualitative evidence of our aggregation strategy's broadens the semantic space of supervision: the t-SNE visualization~\citep{tsne} of sentence embeddings~\citep{sentencebert} from sampled captions reveals distinct yet complementary clustering patterns by source that demonstrate complementary linguistic perspectives.
AudioSetCaps and WavCaps occupy nearby regions associated with general sound-event descriptions and remain close to the human-annotated datasets, whereas JamendoMaxCaps forms a more distinct cluster centered on music-specific terminology and ParaSpeechCaps emphasizes speaking style and paralinguistic attributes.
These patterns suggest that the constituent datasets contribute different descriptive perspectives on audio, thereby broadening what aspects of audio are described.

Quantitative analysis reveals a more nuanced picture (Table~\ref{tab:lexical_stat}). 
CaptionStew substantially expands vocabulary size (56,586 unique words vs. 4,060-27,906 for individual datasets)
However, this growth does not translate into proportionally strong lexical diversity.
The Distinct-n ~\citep{distinct-n} remain low relative to image caption dataset~\citep{cc12m} and text corpora~\citep{wikitext103}, with especially limited surface-form variation in datasets such as JamendoMaxCaps and ParaSpeechCaps. 
In other words, aggregation improves semantic coverage more clearly than it improves lexical diversity.

These findings highlight that simply combining datasets doesn't guarantee improved lexical diversity, revealing broader limitations in current ALP approaches.
Also, the constrained diversity in certain aspect may partially explain weaker scaling behavior observed for certain tasks, as models encounter repetitive linguistic patterns despite increased data volume, aligning with vision-language findings on caption diversity's importance for representation quality~\citep{captionworth,whatincaption}.
This analysis motivates developing enhanced aggregation pipeline and more diverse caption generation methods to better capture the full spectrum of information in audio signals, thereby fully realizing the potential of large-scale ALP.

\section{Conclusions}
We revisited audio–language pretraining with the goal of establishing a rigorous baseline for general-purpose audio representation learning. By aggregating and harmonizing diverse datasets into CaptionStew, we addressed the data scarcity issues that have hindered the field and enabled a rigorous comparison of training objectives and data scales. Our comprehensive evaluation yielded several actionable insights: (1) audio–language pretraining produces competitive representations across speech, music, and environmental sounds; (2) contrastive and captioning objectives exhibit complementary strengths regarding efficiency and scalability; and (3) standard supervised initializations may be unnecessary or even detrimental at scale. Finally, our analysis highlighted the limited lexical diversity in current caption datasets as a key frontier for future improvement. We hope these empirical foundations will accelerate the development of future general-purpose audio representation learning.

\section*{Limitation}
While this work provides valuable empirical insights for audio-language pretraining, we acknowledge several important limitations that present opportunities for future research.

\noindent\textbf{Dataset Construction and Quality.} CaptionStew aggregates captions from multiple sources with varying generation methodologies, including LLM-synthesized descriptions that may introduce systematic biases or artifacts. 
We do not perform extensive quality control or human verification across the aggregated corpus, which could impact model training.
Additionally, our dataset analysis reveals that simple aggregation does not guarantee improved linguistic diversity—CaptionStew's lexical diversity metrics remain lower than mature image-text corpora. 
However, our design choice prioritizes semantic diversity over linguistic variety, as evidenced by the t-SNE clustering analysis showing distinct descriptive focuses across constituent datasets.
While more sophisticated curation strategies could improve quality, our goal was to establish whether diverse caption aggregation can benefit audio representation learning, which our results support despite these limitations.

\noindent\textbf{Limited Technical Novelty.} Our work primarily combines existing techniques—contrastive learning, captioning objectives, and dataset aggregation—rather than introducing fundamentally new methods. The mixed autoregressive/parallel training approach is adapted from vision-language work (CapPa), and our architectural choices follow standard practices.
We acknowledge that the technical contributions are largely empirical rather than methodological.
However, this aligns with our primary goal of systematically evaluating audio-language pretraining's potential for general-purpose representation learning. 
The field currently lacks comprehensive comparative studies across objectives, evaluation protocols, and training factors.
Our systematic analysis reveals important insights about scaling behaviors and initialization effects that have practical implications for practitioners, even if the underlying techniques are not novel.

\noindent\textbf{Limited Model and Data Scalability.} Our experiments are constrained to 10M audio-text pairs and relatively modest model sizes compared to state-of-the-art vision-language systems that leverage billions of samples and much larger architectures. 
This scale limitation may not fully reflect the potential of audio-language pretraining, particularly for the captioning objective which our results suggest benefits from larger-scale training. 
Additionally, we do not explore recent advances in large language model integration or more sophisticated architectural designs that could improve performance. These constraints stem from computational resource limitations and our focus on controlled comparisons rather than pushing absolute performance boundaries. Future work with larger scales may reveal different scaling dynamics and stronger evidence for general-purpose capabilities.

\section*{Ethical Considerations}
This work investigates audio–language pretraining (ALP) as a framework for learning general-purpose audio representations through large-scale empirical analysis. While our study is methodological in nature and does not directly deploy end-user systems, it raises several ethical considerations related to data sources and potential misuse.

\noindent\textbf{Data sourcing and privacy.}
CaptionStew is constructed by aggregating existing open-source audio–text datasets. These datasets are collected under diverse licenses and data collection practices, and may include audio containing human speech or environmental recordings. We rely on the original dataset providers’ compliance with consent, anonymization, and licensing requirements, and we do not introduce new data collection or re-identification procedures. Nevertheless, large-scale aggregation may amplify latent biases or artifacts present in individual sources, including demographic imbalance, recording context skew, or stylistic biases introduced by caption generation process. Users of the dataset and pretrained models should be aware of these limitations.

\noindent\textbf{Potential misuse.}
General-purpose audio representations can benifit applications such as accessibility tools, audio search, and audio understanding, but they may also lower barriers to harmful uses, including surveillance, profiling, or misuse of speaker-related attributes. Our released models are intended for research purposes, and we do not claim suitability for high-stakes or safety-critical scenario without safeguards and validation.

In summary, this work aims to clarify the capabilities and limitations of audio–language pretraining through transparent empirical study. We hope to support more responsible development and evaluation of general-purpose audio representation learning.
\newpage
\bibliography{custom}

\newpage
\appendix
\section{Appendix}
\label{sec:appendix}

\begin{figure*}[h!]
    \centering
    \includegraphics[width=0.95\linewidth]{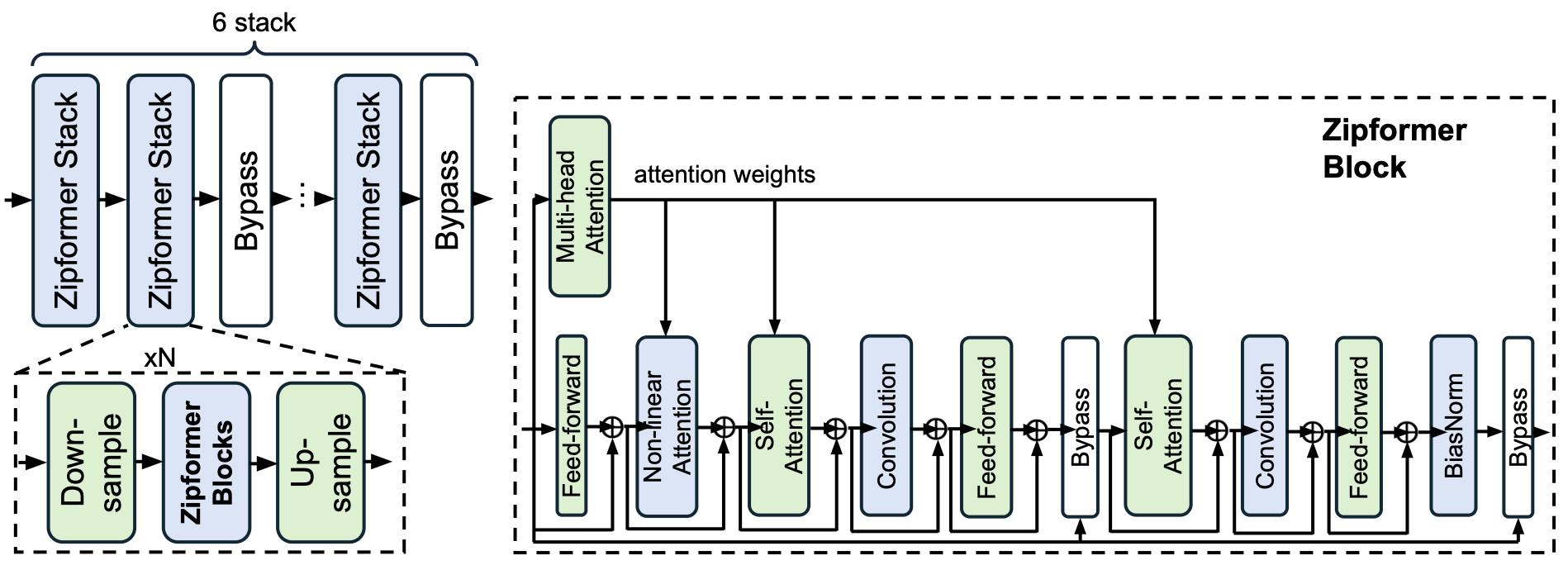}
    \caption{Model diagram of Zipformer.}
    \label{fig:zipformer}
\end{figure*}

\begin{table*}
\caption{Zipformer performance across audio domains when trained from scratch on individual datasets, demonstrating cross-domain efficacy as a general audio encoder.}
\label{table:zipformer}
\centering
\resizebox{0.95\textwidth}{!}{
\begin{tabular}{ccccccc}
\toprule
AudioSet (mAP) & VggSound (acc) & VoxCeleb2 (acc) & CREMA (acc) & MagnaTagATune (mAP) & NSynth-Instrument (acc) \\
\midrule
0.46           & 54.2           & 84.8            & 65.4        & 0.38                & 78.8             \\
\bottomrule
\end{tabular}
}
\end{table*}
\subsection{Full Implementation details}
\label{app:full_implementation_details}
We pretrain all models on CaptionStew.
Training data preparation is performed with the Lhotse~\citep{zelasko2021lhotse} toolkit. 
All audio is resampled to 16 kHz and converted into 80-dimensional log-Mel filterbank features using a 25 ms window length and 10 ms hop size. Text is tokenized with a 50k-vocabulary BPE tokenizer~\citep{lewis2019bart}.

The audio encoder uses a Zipformer-M architecture (see Appendix~\ref{app:zipformer}), chosen for its efficiency on long sequences and fast convergence.
For contrastive pretraining, the text encoder follows BERT-base architecture (12 layers 768 hidden dimensions)~\citep{devlin2019bert}.
For captioning pretraining, the text decoder adopts the BART-base decoder architecture (6 layers, 768 hidden dimensions)~\citep{lewis2019bart}. 
For the decoding mode, the ratio between autoregressive and parallel decoding is 0.25:0.75. 
It is worth noting that we use twice as many encoder layers as decoder layers to ensure comparable training speed across objectives.

Following prior works in audio-language pretraining~\citep{clap, LAION-CLAP, wavcaps,audiosetcaps}, we experiment with two scenarios: training from scratch (denoted by \textit{-scratch}) or initialized from pretrained checkpoints (denoted by \textit{-init}).
The audio encoder initializes from a Zipformer-based audio event classifier (Zipformer-AEC) trained on AudioSet~\citep{AudioSet} with an mAP of 0.46, while text components use corresponding publicly available checkpoints.
All models are trained on 8 Tesla V100 GPUs with an effective batch size of 640 seconds of audio per GPU.
Training runs for 600k steps from scratch (14 days wall-clock time) or 200k steps if initialized from pretrained checkpoint.

\newpage
\subsection{Zipformer Model}
\label{app:zipformer}
Zipformer (Figure~\ref{fig:zipformer}) employs a U-Net-inspired design with six Transformer stages that process sequences at multiple temporal resolutions. The stages operate at progressively decreasing then increasing frame rates (50, 25, 12.5, 6.25, 12.5, and 25 Hz), with residual and upsampling connections between stages to capture both fine-grained and long-range temporal patterns. We implement the original \{2,2,3,4,3,2\} block configuration, where each number indicates the blocks per stage. After processing through all stages, outputs are fused at 25 Hz to produce frame-level embeddings. The model incorporates several architectural improvements: BiasNorm for gradient stability, Swoosh activation functions for better convergence, and ScaledAdam optimizer. The resulting embeddings are 768-dimensional and used consistently across all downstream evaluation tasks.

Although Zipformer was originally designed for automatic speech recognition, we conducted preliminary experiments to validate its effectiveness as a general audio encoder across diverse domains.
As in Table~\ref{table:zipformer}, our initial studies confirmed that Zipformer achieves competitive performance on environmental sound classification, music understanding, and speaker-related tasks, demonstrating its suitability as a unified backbone for multi-domain audio representation learning. This cross-domain efficacy makes it an appropriate choice for our experiments.

\begin{table*}[!h]
\renewcommand{\arraystretch}{0.9}
\caption{Details of public-available datasets contribute to proposed CaptionStew dataset. We summarize their size, domain coverage, audio sources, captioning style, and generation pipelines.}
\label{tab:source}
\resizebox{\textwidth}{!}{
\begin{tabular}{llllll}
\toprule
\textbf{Dataset}                                                      & \textbf{\#audio/\#cap}                                                                 & \textbf{Domain}                                       & \textbf{Audio source}                                                                                 & \textbf{Caption style}                                                                                                                                      & \textbf{Caption generation pipeline}                                                                                                                          \\
\midrule
\begin{tabular}[c]{@{}l@{}}AudioCaps\\ \citep{audiocaps}\end{tabular} & 46k/46k                                                                       & \begin{tabular}[c]{@{}l@{}}general (environmental,\\  human/animal sounds)\end{tabular} & \begin{tabular}[c]{@{}l@{}}AudioSet\\ \citep{AudioSet}\end{tabular}                                                                                     & \begin{tabular}[c]{@{}l@{}}Human-annotated, short description\end{tabular}                                                                                                        & crowdsourced                                                                                                                                         \\[1.0em] \midrule
\begin{tabular}[c]{@{}l@{}}Clotho\\ \citep{clotho}\end{tabular}                                                       & 5k/25k                                                                        & environmental sounds                         & FreeSound                                                                                    & \begin{tabular}[c]{@{}l@{}}Human-annotated, short description\end{tabular}                                                                                       & crowdsourced                                                                                                                                         \\[1.0em] \midrule
\begin{tabular}[c]{@{}l@{}}MusicCaps\\ \citep{MusicCaps}\end{tabular}                                                   & 3k/3k                                                                         & music                                        & AudioSet                                                                         & \begin{tabular}[c]{@{}l@{}}Expert musician-written,\\multi-sentence, fine-grained description\end{tabular}                                                                                  & expert curation                                                                                                                                      \\[1.0em] \midrule
\begin{tabular}[c]{@{}l@{}}WavCaps\\ \citep{wavcaps}\end{tabular}                                                       & 400k/400k                                                                     & \begin{tabular}[c]{@{}l@{}}general (environmental,\\  human/animal sounds)\end{tabular} & \begin{tabular}[c]{@{}l@{}} AudioSet\\ BBC Sound Effect\\ FreeSound\\ SoundBible\end{tabular} & LLM-refined captions                                                                                 & \begin{tabular}[c]{@{}l@{}}three-stage pipeline:\\web-crawled raw descriptions\\ $\rightarrow$ ChatGPT rewrite $\rightarrow$ filtering\end{tabular} \\[1.0em] \midrule
\begin{tabular}[c]{@{}l@{}}AudioSetCaps\\ \citep{audiosetcaps}\end{tabular}                                                  & \begin{tabular}[c]{@{}l@{}}1.9M/1.9M\\ 4.0M/4.0M\\ 182k/182k\end{tabular}     & \begin{tabular}[c]{@{}l@{}}general (environmental,\\  human/animal sounds)\end{tabular} & \begin{tabular}[c]{@{}l@{}}AudioSet\\ YouTube8M\\\citep{youtube8m}\\ VggSound\\\citep{vggsound}\end{tabular}                      & \begin{tabular}[c]{@{}l@{}}LLM-generated, detailed,\\multi-sentence description \end{tabular}                                                                                                   & \begin{tabular}[c]{@{}l@{}}three-stage pipeline:\\LALM attribute extraction \\ $\rightarrow$ LLM captioning \\$\rightarrow$ CLAP-based filtering\end{tabular}   \\[1.0em] \midrule
\begin{tabular}[c]{@{}l@{}}FusionAudio\\ \citep{fusionaudio}\end{tabular}                                    & 1.2M/1.2M                                                                     & \begin{tabular}[c]{@{}l@{}}general (environmental,\\  human/animal sounds)\end{tabular} & AudioSet       & \begin{tabular}[c]{@{}l@{}}LLM-augmented, multi-sentence,\\visual-enhanced description \end{tabular}                                                                                                    & \begin{tabular}[c]{@{}l@{}}multimodal context fusion \\(audio, visual, metadata)\\ + LLM captioning\end{tabular}                                       \\[1.0em] \midrule
\begin{tabular}[c]{@{}l@{}}JamendoMaxCap\\ \citep{jamendomaxcaps}\end{tabular}           & 360k/1.8M                                                                     & music                                        & Jamendo Platform                                                                                     & \begin{tabular}[c]{@{}l@{}}LLM-augmented, multi-sentence,\\fine-grained music description \end{tabular}                                                                                              & \begin{tabular}[c]{@{}l@{}}retrieval-based\\metadata imputation\\ + LLM captioning\end{tabular}                                                       \\[1.0em] \midrule
\begin{tabular}[c]{@{}l@{}}ParaSpeechCaps\\ \citep{paraspeechcaps}\end{tabular}             & \begin{tabular}[c]{@{}l@{}}116k/116k (base)\\ 924k/924k (scaled)\end{tabular} & expressive speech                            & \begin{tabular}[c]{@{}l@{}}VoxCeleb1\\\citep{voxceleb1}\\ VoxCeleb2\\\citep{voxceleb2}\\ EARS\\\citep{ears}\\ Expresso\\\citep{nguyen2023expresso}\\ Emilia\\\citep{he2024emilia}\end{tabular}     & \begin{tabular}[c]{@{}l@{}}Human-annotated/LLM-augmented,\\speaking-style description\end{tabular} & \begin{tabular}[c]{@{}l@{}}crowdsourced / \\ retrieval-based\\metadata imputation\\ + LALM captioning\end{tabular}    \\
\bottomrule
\end{tabular}
}
\end{table*}
\begin{table*}[!h]
\caption{Example caption sampled from each sourced dataset.}
\label{tab:example}
\resizebox{\textwidth}{!}{
\begin{tabular}{ll}
\toprule
\textbf{Dataset}        & \textbf{Example Caption }                                                                                                                                                                                                                                                                                                                                                                                                                                                                  \\ \midrule
AudioCaps      & "Distant traffic sounds followed by a car passing closely."                                                                                                                                                                                                                                                                                                                                                                                                                       \\ \midrule
Clotho         & "Something is being sanded or dragged, manipulated, scraped."                                                                                                                                                                                                                                                                                                                                                                                                                     \\ \midrule
MusicCaps      & \begin{tabular}[c]{@{}l@{}}"This is an advertisement jingle music piece. It is an instrumental piece. The main theme is being played by the piano while \\ there is a synth string sound in the melodic background. There is an emotional, heart-touching atmosphere. This piece could \\ be used in the soundtrack of a drama movie during scenes of tragedy. It could also work well as an advertisement jingle \\ where there is an attempted appeal to emotion."\end{tabular} \\
WavCaps        & "Music is playing while people are walking and crickets are chirping."                                                                                                                                                                                                                                                                                                                                                                                                            \\ \midrule
AudioSetCaps   & \begin{tabular}[c]{@{}l@{}}"A choir performs a folk music piece, utilizing only their voices as instruments. The harmonious and uplifting sounds create\\ an engaging and captivating listening experience."\end{tabular}                                                                                                                                                                                                                                                         \\ \midrule
FusionAudio    & "A full choir is singing with powerful harmonized vocals"                                                                                                                                                                                                                                                                                                                                                                                                                         \\ \midrule
JamendoMaxCaps & \begin{tabular}[c]{@{}l@{}}"The music is instrumental with a dominant piano sound, falling under the genres of ambient, classical, and contemporary. \\ It carries a mood that is nostalgic and romantic, played in a 4/4 time signature at a tempo of 81.1 bpm. The piano piece evokes\\  a sense of tranquility, making it suitable for scenarios depicting love scenes or peaceful moments in movies."\end{tabular}                                                            \\ \midrule
ParaSpeechCaps & \begin{tabular}[c]{@{}l@{}}"A male speaker delivers his words quickly with a medium-pitched voice. His speech exhibits a flowing rhythm and is recorded\\  in an environment that is balanced in clarity. There is a subtle nasal quality to his speech, suggesting an American accent."\end{tabular}     \\
\bottomrule
\end{tabular}
}
\end{table*}

\newpage
\subsection{Sourced Datasets for CaptionStew}
\label{app:dataset_detail}
CaptionStew aggregates eight open-source audio caption datasets to address data scarcity and limited diversity in current audio-language pretraining. The constituent datasets span environmental sounds, music, and expressive speech, with fundamentally different captioning approaches—from crowdsourced human annotation to expert curation to various LLM-based generation pipelines.
Table~\ref{tab:source} and Table~\ref{tab:example} detail each dataset's characteristics and provide example captions that illustrate the diverse descriptive styles, ranging from concise event descriptions to detailed multi-sentence narratives with fine-grained acoustic and contextual information.
During aggregation, we filter audio samples longer than one minute for computational efficiency and remove samples that overlap with common audio understanding benchmarks~\citep{audiocaps,clotho,audiocaps,MusicCaps,fsd50k,vggsound, US8K} to prevent data leakage. This approach preserves the unique characteristics of each source while creating a unified corpus that captures broader semantic coverage than individual datasets.
Moreover, when identical audio clips appear across source datasets, we consolidate them into a single audio instance and merge their associated captions into a multi-caption set, rather than treating duplicated waveforms as separate training examples. This prevents overlapping recordings from being over-counted while still preserving complementary textual supervision for the same audio. We note that the resulting mixture remains heterogeneous and imbalanced across sources, reflecting the current landscape of publicly available audio–caption data; our goal is not to enforce a perfectly balanced corpus, but to provide a realistic large-scale testbed for studying audio–language pretraining under diverse supervision.

\begin{table*}[!h]
    \centering
    \caption{Details of the dataset used for assessing audio representation. $^\dagger$evaluate by GPT-4 in AIR-Bench. $^\ddagger$synthesized with public available speech datasets~\citep{ardila2019common, busso2008iemocap, cao2014crema, ravdess, poria2018meld} with fixed question template.}
    \label{tab:dataset_full}
    \renewcommand{\arraystretch}{0.86}
    \resizebox{1.0\textwidth}{!}{
\begin{tabular}{llccccl}
\toprule  
Evaluation Dataset & Task   & \#samples & \#class & train  & eval                  & Metrics \\
\midrule
FSD-50k            & Multi-label audio event classification & 37,168 / 10,231 & 200 & $\checkmark$ & $\checkmark$ & mAP     \\
VggSound           & Single-label audio event classification & 183,730 / 15,446 & 309 & $\checkmark$ & $\checkmark$ & accuracy    \\
VoxCeleb2          & Speaker identification        & 1,092,009 / 36,693 & 5,994 & $\checkmark$ & $\checkmark$           & accuracy    \\
CREMA-D            & Speech emotion recognition    & 6,030 / 706 & 6 & $\checkmark$ & $\checkmark$                   & accuracy   \\
MagnaTagATune      & Music tagging           & 19,425 / 4,856 & 50 & $\checkmark$ & $\checkmark$                            & mAP     \\
NSynth             & Musical instrument classification   & 289,205 / 4,096 & 11 & $\checkmark$ & $\checkmark$                       & accuracy  \\
AudioSet-strong    & Sound event detection & 103,463 / 16,996 & 456 & $\checkmark$ & $\checkmark$     & PSDS1 \\
\midrule
AudioCaps             & \multirow{3}{*}{\begin{tabular}[c]{@{}l@{}}Text-to-audio retrieval\\ Audio captioning\end{tabular}}  & 49,838 / 975 & -- & $\checkmark$ & $\checkmark$     & \multirow{3}{*}{\begin{tabular}[c]{@{}l@{}}Recall@1\\ RougeL\end{tabular}}   \\
ParaSpeechCaps          &   &    116,516 / 500 & -- & $\checkmark$ & $\checkmark$          &    \\
MusicCaps             &    &    2,663 / 500 & -- & $\checkmark$ & $\checkmark$            &    \\
\midrule
ClothoAQA & \multirow{8}{*}{Open-formed question answering} &    7,044 & -- & $\checkmark$ & $\times$          & \multirow{8}{*}{Score$^\dagger$}\\
ParaLMQA$\ddagger$ & &  160,000 & -- & $\checkmark$ & $\times$       & \\
MusicQA & & 70,011 & -- & $\checkmark$ & $\times$          & \\
AIRBench-chat-sound & &  400 & -- & $\times$ & $\checkmark$       & \\
AIRBench-foundation-emotion & &  1,000 & -- & $\times$ & $\checkmark$       & \\
AIRBench-foundation-gender & &  1,000 & -- & $\times$ & $\checkmark$       & \\
AIRBench-foundation-age & &  1,000 & -- & $\times$ & $\checkmark$       & \\
AIRBench-chat-sound & &  400 & -- & $\times$ & $\checkmark$      & \\
\bottomrule
\end{tabular}
}
    
    \label{tab:placeholder}
\end{table*}

\subsection{Evaluation Datasets}
\label{app:eval_datasets}
Table~\ref{tab:dataset_full} details the evaluation datasets and their metrics used for assessing audio representation quality across our three evaluation protocols: linear probing~\cite{fsd50k, vggsound, voxceleb2, cao2014crema, MTT, NSynth, asstrong, ebbers2022threshold}, audio-language alignment~\cite{audiocaps, paraspeechcaps, MusicCaps, lin-2004-rouge} and open-form question answering~\cite{clothoaqa, mullama, huo2025auden, airbench}.

\subsection{Main Results (cont.)}
Table 9 presents linear probing results when using multi-head attention pooling instead of mean pooling. 
With learned attention pooling, the performance gap between contrastive and captioning objectives narrows substantially, particularly evident on speaker identification where captioning-scratch achieves 72.86\% compared to 46.67\% with mean pooling (Table~\ref{tab:main}).
This demonstrates that captioning models benefit significantly from adaptive pooling mechanisms, while contrastive learning's explicit optimization for clip-level representations shows less sensitivity to pooling strategy.
These results underscore the critical importance of appropriate downstream module selection when evaluating different pretraining paradigms, as the choice of pooling mechanism can dramatically influence conclusions about objective effectiveness.
The improved performance across all methods with attention pooling also suggests that frame-level representations from both objectives contain rich information that can be better exploited through learned aggregation. SOTA results and SSL baseline results in Table~\ref{tab:main} and Table~\ref{tab:appendix_main} are quoted collectively from ~\citet{niizumi2025m2d2, hear, li2022atst, HARES, bharadwaj2025openbeats, gong2022contrastive, lanzendorfer2025bootstrapping, audiosetcaps, airbench}.
\label{app:main}
\begin{table*}[!h]
\caption{Linear probing results when using multi-head attention pooling.}
\label{tab:appendix_main}
\resizebox{\textwidth}{!}{%
\begin{tabular}{lcccccccc}
\toprule
\multirow{3}{*}{Method} & \multirow{3}{*}{\shortstack{Model\\Initialization}}& \multirow{3}{*}{\shortstack{Audio-language\\Pretraining}}& \multicolumn{6}{c}{\textbf{linear probing}}                                    \\
\cmidrule{4-9}
                  & & & \twolineheader{AEC}{FSD50k} & \twolineheader{AEC}{VggSound} & \twolineheader{SID}{VoxCeleb2} & \twolineheader{SER}{CREMA} & \twolineheader{MTAG}{MagnaTagATune} & \twolineheader{INST}{NSynth} \\
\midrule
\textbf{\textit{   Our Supervised Baselines}} & & & & \\
Zipformer-AEC~\cite{Zipformer}        & AS SL  &  --  & 0.656   & \underline{56.23}    & 58.76     & 72.52 & 0.405   & 67.19   \\
\midrule
\textbf{\textit{   Our Audio-language Pretrained Models}} & & & & \\
Contrastive-\textit{scratch}       &     --     & CS10M & 0.640   & 52.81    & \textbf{72.86}     & \underline{\textbf{74.50}} & 0.406   & 75.00  \\
Captioning-\textit{scratch}        &    --      & CS10M & 0.619   & 50.97    & 56.64     & 70.40 & 0.406   & 72.10   \\
Contrastive-\textit{init}       & AS SL      & CS10M & \underline{\textbf{0.670}}   & \textbf{54.89}    & 72.24     & 73.09 & \underline{\textbf{0.412}}   & \textbf{76.70} \\
Captioning-\textit{init}        & AS SL      & CS10M & 0.660   & 53.68    & 62.24     & 71.67 & 0.411   & 74.49  \\
\midrule
\midrule
SOTA$^\ddagger$ & & & 0.655 & 59.50 & 96.20 & -- & 0.414 & 79.20 \\
\bottomrule
\end{tabular}%
}
\end{table*}

\subsection{Additional Results}
Aside from learning representations, we also compare against state-of-the-art audio-text retrieval models to assess our approach's performance on the specific task it was designed for. Table~\ref{table:with_audiosetcaps} presents retrieval results for our best-performing model (Contrastive-init) against state-of-the-art
audio-text retrieval model~\citep{audiosetcaps}. 
Our model achieving comparable or superior results on benchmarks in various audio domains, with particularly strong performance on speech and music retrieval.
The results indicate that our general-purpose audio-language pretraining approach can compete with specialized retrieval models while offering broader applicability across diverse usage scenarios.
\begin{table*}[!h]
\caption{audio-text retrieval of the best performing model (Contrastive-init) against state-of-the-art audio-text retrieval model. $^\dagger$reproduce by ourselves. }
\label{table:with_audiosetcaps}
\resizebox{\textwidth}{!}{
\begin{tabular}{lcccccc}
\toprule
\multirow{2}{*}{Model}  & \multicolumn{3}{c}{Text-to-audio}          & \multicolumn{3}{c}{Audio-to-text}          \\
\cmidrule(lr){2-4} \cmidrule(lr){5-7}
                        & AudioCaps   & ParaSpeechCaps & MusicCaps   & AudioCaps   & ParaSpeechCaps & MusicCaps   \\
                        \midrule
AudioSetCaps$^\dagger$            & 49.7 / 79.2 & 0.8 / 2.5      & 13.4 / 30.6 & 45.9 / 80.8 & 0.2 / 3.8      & 12.0 / 29.0 \\
Contrastive-init (ours) & 44.4 / 79.0 & 29.6 / 61.6    & 22.4 / 53.0 & 47.2 / 78.8 & 27.0 / 57.4    & 26.0 / 56.2 \\
\bottomrule
\end{tabular}
}
\end{table*}


\subsection{The Use of Large Language Model}
The authors used large language models to assist with writing refinement and grammatical corrections during the drafting process. All technical content, experimental design, analysis, and conclusions remain the authors' original contributions.

\begin{table*}
\centering
\caption{Comparison of lexical statistics and diversity across audio caption datasets and text corpora. We report vocabulary size (\#vocab), average sentence length (avg. sent), and Distinct-n.}
    \begin{tabular}{llcccccc}
\toprule
\multirow{2}{*}{Source}  & \multirow{2}{*}{\#vocab} & \multirow{2}{*}{avg. sent} & \multicolumn{4}{c}{Distinct-n} \\
\cmidrule{4-7}
                        &        &                    & 1      & 2     & 3     & 4     \\
\midrule
AudioCaps               & 5,572                    & 8.46                       & 0.011  & 0.113 & 0.309 & 0.519 \\
WavCaps                 & 18,372                   & 7.77                       & 0.026  & 0.184 & 0.420 & 0.646 \\
AudioSetCaps            & 21,061                   & 28.22                      & 0.006  & 0.082 & 0.249 & 0.450 \\
FusionAudio             & 18,403                   & 13.81                      & 0.009  & 0.111 & 0.322 & 0.546 \\
JamendoMaxCaps          & 27,906                   & 63.29                      & 0.002  & 0.026 & 0.079 & 0.153 \\
ParaSpeechCaps          & 4,060                    & 28.50                      & 0.001  & 0.015 & 0.051 & 0.112 \\
\midrule
CaptionStew(Ours)            & 56,586                   & 32.23                      & 0.006  & 0.080 & 0.231 & 0.401 \\
\midrule
CC12M                   & 366,175                  & 17.03                      & 0.046  & 0.486 & 0.813 & 0.927 \\
WikiText-103            & 531,346                  & 74.29                      & 0.031  & 0.365 & 0.757 & 0.930 \\
\bottomrule
\end{tabular}
\
\label{tab:lexical_stat}
\end{table*}

\newpage
\begin{figure*}
    \centering
    \includegraphics[width=0.85\linewidth]{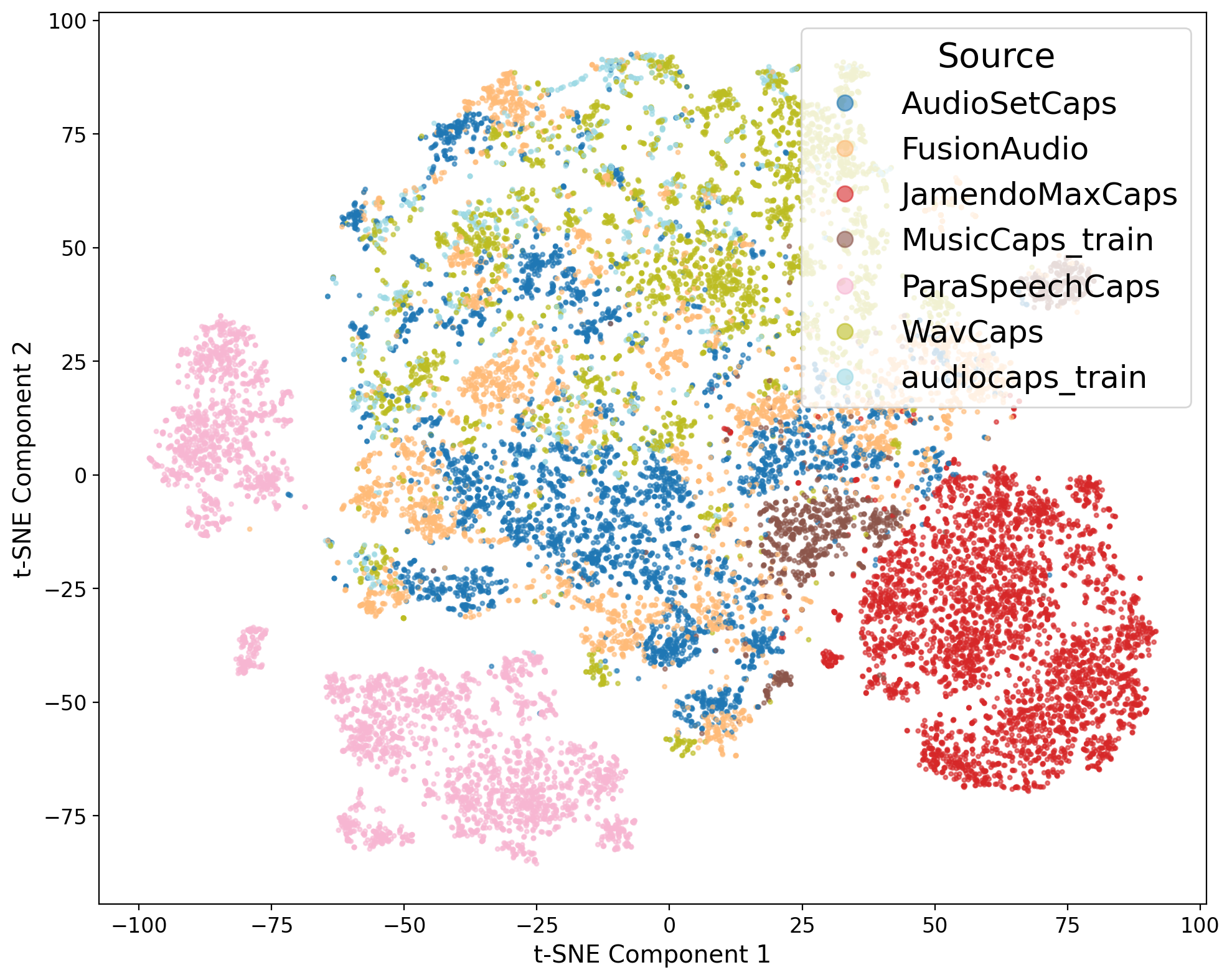}
    \caption{t-SNE visualization of sentence embedding of captions grouped by source.}
    \label{fig:tsne}
\end{figure*}


\end{document}